\documentclass[aps,prl,groupedaddress,fleqn,twocolumn]{revtex4}
\pdfoutput=1
\usepackage{graphicx}
\usepackage{xcolor}
\usepackage{ulem}
\usepackage{amsmath}
\usepackage{gensymb}
\usepackage{float}
\usepackage{physics}
\usepackage{amsfonts}
\newcommand{\comment}[1]{}
\begin{document}
\title{Universal interrelation between dynamics and thermodynamics and a dynamically-driven ``c''-transition in fluids}
\author{C. Cockrell$^{1,*}$, V. V. Brazhkin$^2$ and K. Trachenko$^1$}
\address{$^1$ School of Physics and Astronomy, Queen Mary University of London, Mile End Road, London, E1 4NS, UK}
\address{$2$ Institute for High Pressure Physics, RAS, 108840, Troitsk, Moscow, Russia}
\address{$*$ Corresponding author: c.j.cockrell@qmul.ac.uk}


\begin{abstract}
Our first very wide survey of the supercritical phase diagram and its key properties reveals a universal interrelation between dynamics and thermodynamics and an unambiguous transition between liquidlike and gaslike states. This is seen in the master plot showing a collapse of the data representing the dependence of specific heat on key dynamical parameters in the system for many different paths on the phase diagram. As a result, the observed transition is path-independent. We call it a ``c''-transition due to the ``c''-shaped curve parameterizing the dependence of the specific heat on key dynamical parameters. The ``c''-transition has a fixed inversion point and provides a new structure to the phase diagram, operating deep in the supercritical state (up to at least 2000 times the critical pressure and 50 times the critical temperature). The data collapse and path independence as well as the existence of a special inversion point on the phase diagram are indicative of either of a sharp crossover or a new phase transition in the deeply supercritical state. 
\end{abstract}

\maketitle

\section{Introduction}

The basic phase diagram of matter charts the areas where solid, liquid and gas exist as physically distinct phases separated by transition lines. The three transition lines all emerge from the triple point, but two of three are finite in length. The sublimation line terminates at zero temperature or pressure, and the boiling line terminates at the critical point. The third line, the melting line, extends to arbitrary temperatures and pressures, so long as the system remains chemically unaltered. A  large area of the phase diagram lies above the critical point, representing the supercritical state of matter.

Until fairly recently, there was no reason to survey the state of matter well above the critical point in any detail. This part of the phase diagram was thought to be physically homogeneous with no discernible differences between liquid-like and gas-like states \cite{landau,deben}. Not far above the critical point, persisting critical anomalies can continue to conditionally separate liquid-like and gas-like states. More recently, experiments and theory have given indications that the entire supercritical state may in fact be inhomonegenous and have states with qualitatively different properties.

Going deeply supercritical in the experiment is often unworkable and, when possible, is limiting in terms of measuring key system properties \cite{nist}. We are much less limited in molecular dynamics (MD) simulations. 

In this paper, we conduct the widest and most detailed survey of the supercritical phase diagram undertaken so far: we extend the parameter range from the melting point and up to 330$T_c$ and 8000$P_c$ ($T_c$ and $P_c$ are critical temperature and pressure for Argon) and traverse it along different isobars, isochores and isotherms. We calculate key thermodynamic and dynamical properties along different paths on the phase diagram. We then show that all data collapse on a single dynamic-thermodynamic ``c''-shaped master curve with a fixed inversion point. The ``c''-transition provides an unambiguous separation of liquidlike and gaslike fluids, extending deep into the supercritical state and therefore challenging the existing view of this state as homogeneous and lacking qualitative transitions. The data collapse and the existence of a special inversion point on the phase diagram are consistent with either a sharp crossover or a new phase transition (perhaps of higher-order) operating in the supercritical state of matter. 

\section{Simulation details}

We simulate a commonly used and well-characterised Lennard-Jones system describing Argon ($\sigma=3.4~\mathrm{\AA}$, $\epsilon = 0.01032$ eV) along three isobars, isotherms, and isochores in the deep supercritical state, plus an additional isobar near the critical point. The supercritical paths are plotted in Fig. \ref{fig:phasediagram}. Our largest temperature and pressure, 330$T_c$ and 8000$P_c$ (about 50,000 K and 40 GPa) are below those where metallisation and ionization start in Ar. Within the simulated range of state points, the ``c''-transition is seen up to 2000 times the critical pressure and 50 times the critical temperature.

\begin{figure}
\begin{center}
{\scalebox{0.5}{\includegraphics{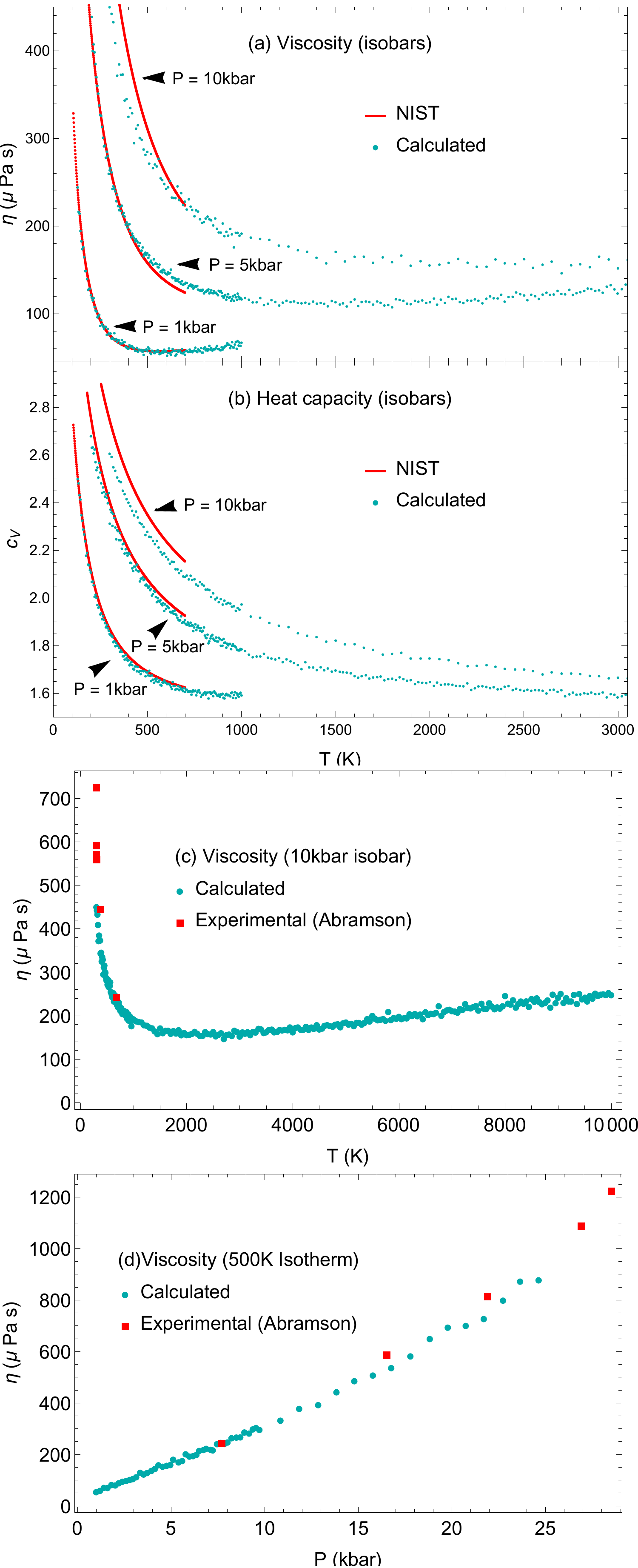}}}
\end{center}
\caption{Comparison of (a) viscosities $\eta$ and (b) isochoric specifc heat capacities $c_V$ ($k_{\rm B}=1$) calculated from simulated trajectories with experimental data from NIST \cite{nist}; comparison of viscosities $\eta$ calculated from simulated trajectories along the (c) 10 kbar isobar and (d) 500 K isotherm, with experimental data from Abramson \cite{abramson}.}
\label{fig:totalcompare}
\end{figure}

Equilibration was performed in the NPT ensemble with the Langevin thermostat in order to generate the mean densities along the isobars and isotherms. System sizes of 500, 4000, and 108,000 atoms were used, with no discrepancy in calculated quantities or our results between these sized. Consistent with the earlier ascertained insensitivity of viscosity to system size \cite{viscositysize}, we find that the viscosity (and the heat capacity) for larger systems of up to 108,000 atoms coincides with that of the smaller systems we simulated. We selected the timestep of 1 fs, which is under 100 times smaller than the fastest oscillations in any system, and conserved total energy under the Velocity-Verlet integrator in the NVE ensemble to one part in $10^5$. Configurations at the target densities on all paths were then generated, which were then equilibrated with the NVT ensemble for 50 ps. Following this equilibration, we generated 20 independent initial conditions for each state point using seeded velocities, and each of these initial conditions were run for 1 ns in the NVE ensemble during which all properties were calculated. We calculated $c_V$ in the NVE ensemble as \cite{Frenkel2001}:

\begin{figure}
\begin{center}
{\scalebox{0.47}{\includegraphics{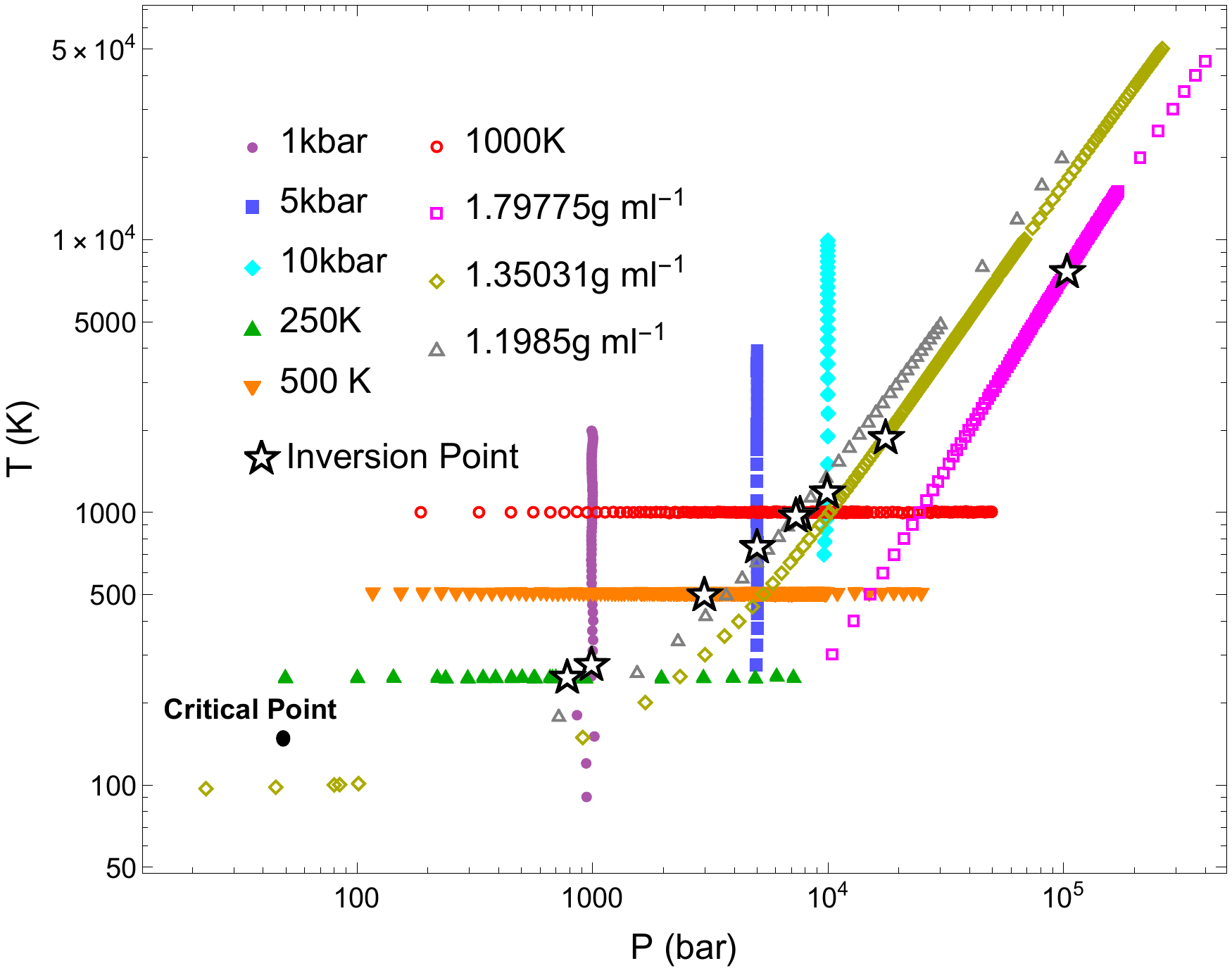}}}
\end{center}
\caption{Paths on the phase diagram and pressure and temperature points explored in this work. Also labelled are the critical point and the state points where the transition at the inversion point at $c_V = 1.9$ takes place.}
\label{fig:phasediagram}
\end{figure}

\begin{equation}
    \label{eqn:cvnve}
    \langle K^2 \rangle - \langle K \rangle^2 = \frac{3}{2} N T^2 \left( 1- \frac{3}{2 c_V}\right),
\end{equation}
with $K$ the total kinetic energy. From here on out, we set $k_{\mathrm{B}} = 1$.  We plot the calculated heat capacities compared to NIST data (where available) \cite{nist} in Fig. \ref{fig:totalcompare}a-b, demonstrating good agreement. The high-frequency shear modulus and shear viscosity were calculated using the molecular stress autocorrelation function, from Green-Kubo theory \cite{Zwanzig1965, Balucani1994}:
\begin{equation}
    \label{eqn:GKshearmod}
    G_\infty = \frac{V}{T} \langle \sigma^{x y}(0)^2 \rangle,
\end{equation}
\begin{equation}
    \label{eqn:GKviscosity}
   \eta = \frac{V}{T} \int_0^{\infty} \rm{d} t \ \langle \sigma^{x y}(t) \sigma^{x y}(0) \rangle,
\end{equation}

\noindent with $\sigma^{x y}$ an off-diagonal component of the microscopic stress tensor. The integration of the long-time tails of autocorrelation functions is a well known issue with the practical implementation of Green-Kubo formulae \cite{Zhang2015}. The 20 independent initial conditions were used to address this, with the autocorrelation function $\langle \sigma^{x y}(t) \sigma^{x y}(0) \rangle$ being averaged over these initial conditions. The end result for viscosity was insensitive to adding more initial conditions and also whether the autocorrelation function was calculated in the NVT or NVE ensembles. 

\section{Results and Discussion}

We zero in on the relationship between thermodynamic and dynamical properties of the supercritical system. The specific heat $c_V$ is an informative choice for a thermodynamic property. The choice of a dynamical property is less obvious because there are several candidates. We first try viscosity, $\eta$, because there is an obvious distinction between its behaviour in the liquid and gas phases. Indeed, $\eta$ decreases and increases with temperature in the liquidlike and gaslike regimes, respectively \cite{sciadv}. In Fig. \ref{fig:totalcompare} we observe a good agreement between calculated and experimental viscosity. This includes good agreement with high-pressure data range \cite{abramson} where the NIST data can be less reliable due to data interpolations used at high pressure.

\begin{figure}
\begin{center}
{\scalebox{0.65}{\includegraphics{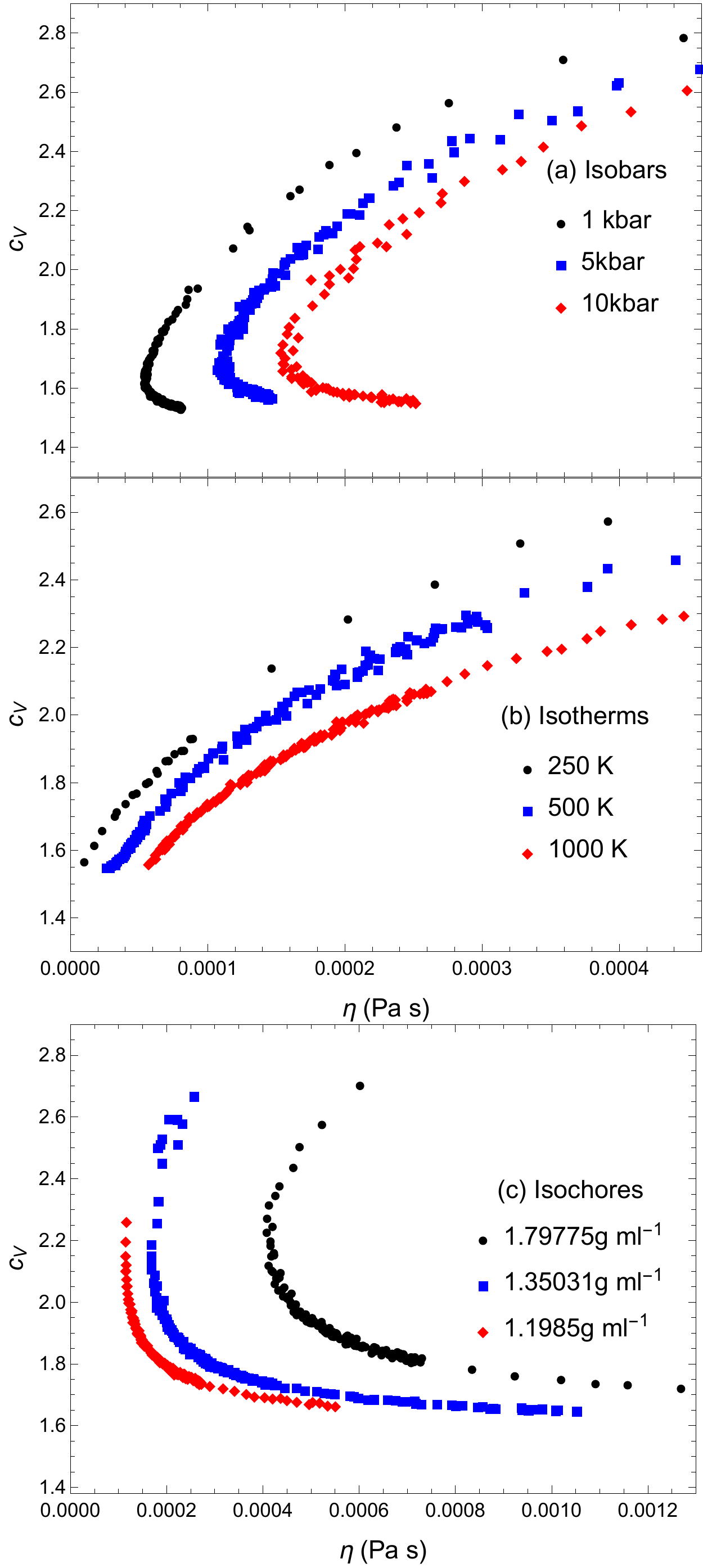}}}
\end{center}
\caption{Specific heat $c_v$ ($k_{\rm B}=1$) as a function of viscosity $\eta$, calculated along three (a) isobars; (b)isotherms; (c) isochores.}
\label{fig:cveta}
\end{figure}

We plot the dependence of $c_V$ on $\eta$ for each of our supercritical phase diagram paths in Fig. \ref{fig:cveta}. Along isobars and isochores, $c_V(\eta)$ has clear turning points due to corresponding minima in $\eta$, whereas no such minima exist along our isothermal paths. We observe strong path dependence in the interrelation between $c_V$ and $\eta$, and will return to the issue of path dependence below. Here we note the earlier work \cite{excess} relating viscosity to excess entropy calculated from the virial expansion at low density. This relation was discussed on empirical grounds and is unrelated to our approach where we discuss the system properties on the basis of excitations that the system possesses. 

Our next important choice for the dynamical parameter is informed by the Maxwell-Frenkel viscoelastic theory \cite{maxwell,frenkel} where the combined elastic and viscous response of a liquid is represented as:

\begin{equation}
\frac{\mathrm{d} s}{\mathrm{d} t}=\frac{\sigma}{\eta}+\frac{1}{G_\infty}\frac{\mathrm{d}\sigma}{\mathrm{d} t}
\label{a1}
\end{equation}

\noindent where $s$ is shear strain, $\eta$ is viscosity, $G$ is high-frequency shear modulus and $P$ is shear stress.

Using \eqref{a1}, Frenkel modified the Navier-Stokes equation to include the elastic response. A simplified form of the resulting equation reads \cite{ropp,physrep}:

\begin{equation}
c^2\frac{\partial^2v}{\partial x^2}=\frac{\partial^2v}{\partial t^2}+\frac{1}{\tau}\frac{\partial v}{\partial t}
\label{gener3}
\end{equation}

\noindent where $v$ is the velocity component perpendicular to $x$, $c$ is transverse wave velocity $c=\sqrt\frac{G}{\rho}$, $\rho$ is density and $\tau=\frac{\eta}{G_\infty}$ is the liquid relaxation time.

Seeking the solution of (\ref{gener3}) as $v=v_0\exp\left(i(kx-\omega t)\right)$ gives $\omega^2+\omega\frac{i}{\tau}-c^2k^2=0$ and $\omega=-\frac{i}{2\tau}\pm\sqrt{c^2k^2-\frac{1}{4\tau^2}}$. If $k<\frac{1}{2c\tau}$, $\omega$ has neither a real part nor transverse propagating modes. For $k>k_g=\frac{1}{2c\tau}$, $v\propto\exp\left(-\frac{t}{2\tau}\right)\exp(i\omega_rt)$, where $\omega_r=\sqrt{c^2k^2-\frac{1}{4\tau^2}}$. Here, 

\begin{equation}
k_g=\frac{1}{2c\tau}
\label{kgap}
\end{equation}

\noindent defines a gapped momentum state seen in several distinct areas of physics \cite{physrep}. In liquids and supercritical fluids, $k_g$ is the important parameter governing the existence of solid-like transverse waves \cite{physrep,prl}.

It follows from the above discussion and Eq. \eqref{kgap} in particular that the solution to the Maxwell-Frenkel viscoelasticity importantly depends on the dynamical fluid elasticity length $\lambda_d=c\tau$ \cite{ropp}. The physical meaning of this length is that it sets the range of propagation of solid-like transverse waves in the liquid because $\tau$ is the time during which the shear stress is relaxed. Hence, $c_V$ should uniquely depend on $\lambda_d$ because (a) $\lambda_d$ governs the propagation of transverse modes and (b) each mode carries the energy $T$ (in harmonic classical case). Therefore, we will use $\lambda_d$ in our subsequent analysis.

\begin{figure}
\begin{center}
{\scalebox{0.6}{\includegraphics{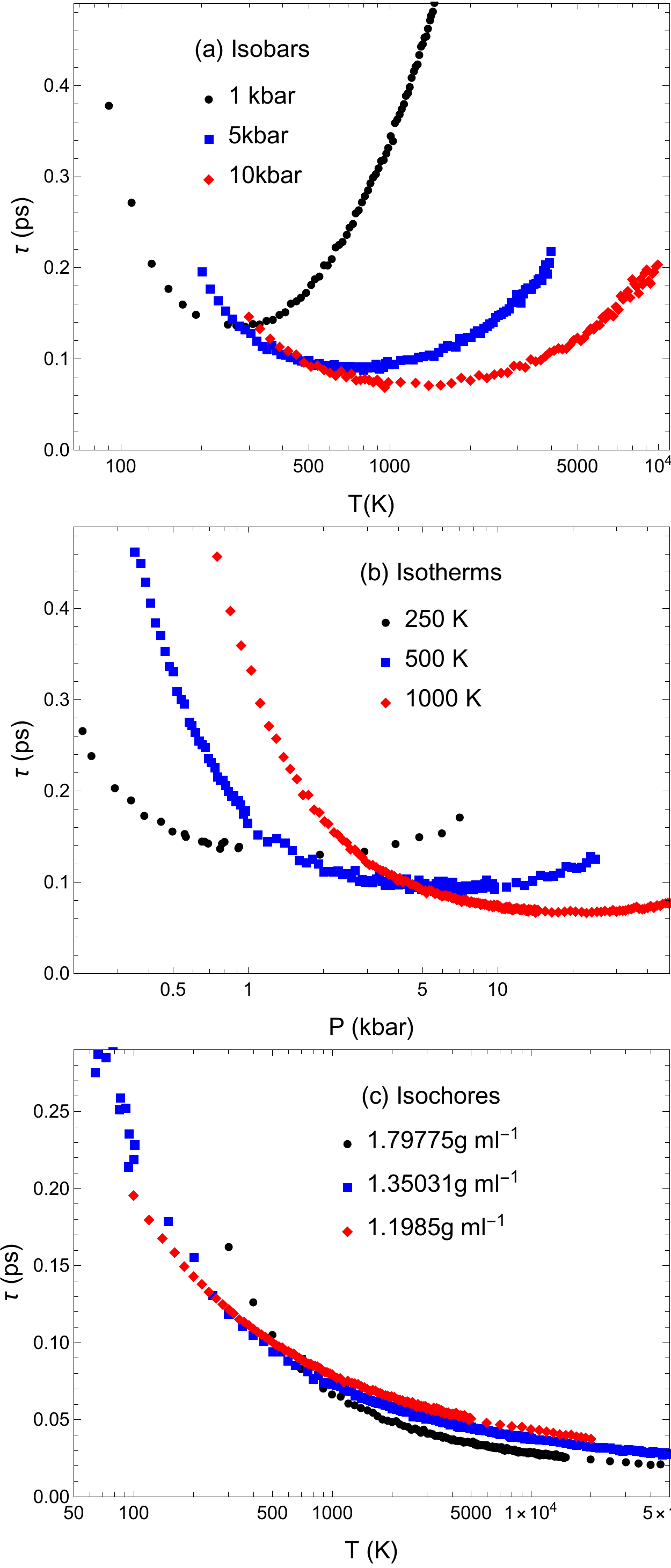}}}
\end{center}
\caption{Liquid relaxation time $\tau$ of molecular dynamics trajectories along three (a) isobars; (b) isotherms; (c) isochores.}
\label{fig:tau}
\end{figure}

Before invoking $\lambda_d$, it is first instructive to look at $\tau(T)$ and $c_v(\tau)$ dependencies. We plot the calculated $\tau=\frac{\eta}{G}$ along our nine supercritical phase diagram paths in Fig. \ref{fig:tau} and observe that $\tau$ has minima along isobars and isotherms and has no minima along isochores. To understand the minima, we recall that Frenkel's theory relates $\tau$ to the average time between molecular rearrangements in the liquid \cite{frenkel}. Backed by experiments and modelling \cite{jac,iwa}, this relation has since become an accepted view \cite{dyre}. In the low-temperature liquid regime, $\tau$ decreases with temperature. In the high-temperature gaslike regime, shear momentum transfer takes place via the collisions of kinetic theory, and relaxation time $\tau$ is therefore interpreted as the mean time between collisions in this regime \cite{frenkel,chapman}. Then, the minima represent a crossover from a liquidlike relaxation time commanded by diffusion and oscillation events \cite{zaccone} to a gaslike relaxation time commanded by collisions \cite{ropp}. This same crossover takes place along isochores, but manifests differently in $\tau$ because $\tau$ can only decrease with temperature on an isochore. Indeed, $\tau=\frac{L}{v_{\rm{th}}}$ in the gaslike regime, where $L$ and $v_{\rm{th}}$ are the particle mean-free path and thermal velocity in the gaslike state and $L=\frac{1}{n A}$, where $n$ is particle density and $A$ is the particle cross-section area. Hence $\tau$ decreases with temperature at constant $n$ mostly because $v_{\rm{th}}\propto\sqrt{T}$ increases ($A$ decreases with temperature weakly). We note that the minima of both $\eta$ and $\tau$ depend on the path taken on the phase diagram and path parameters. 

\begin{figure}
\begin{center}
{\scalebox{0.65}{\includegraphics{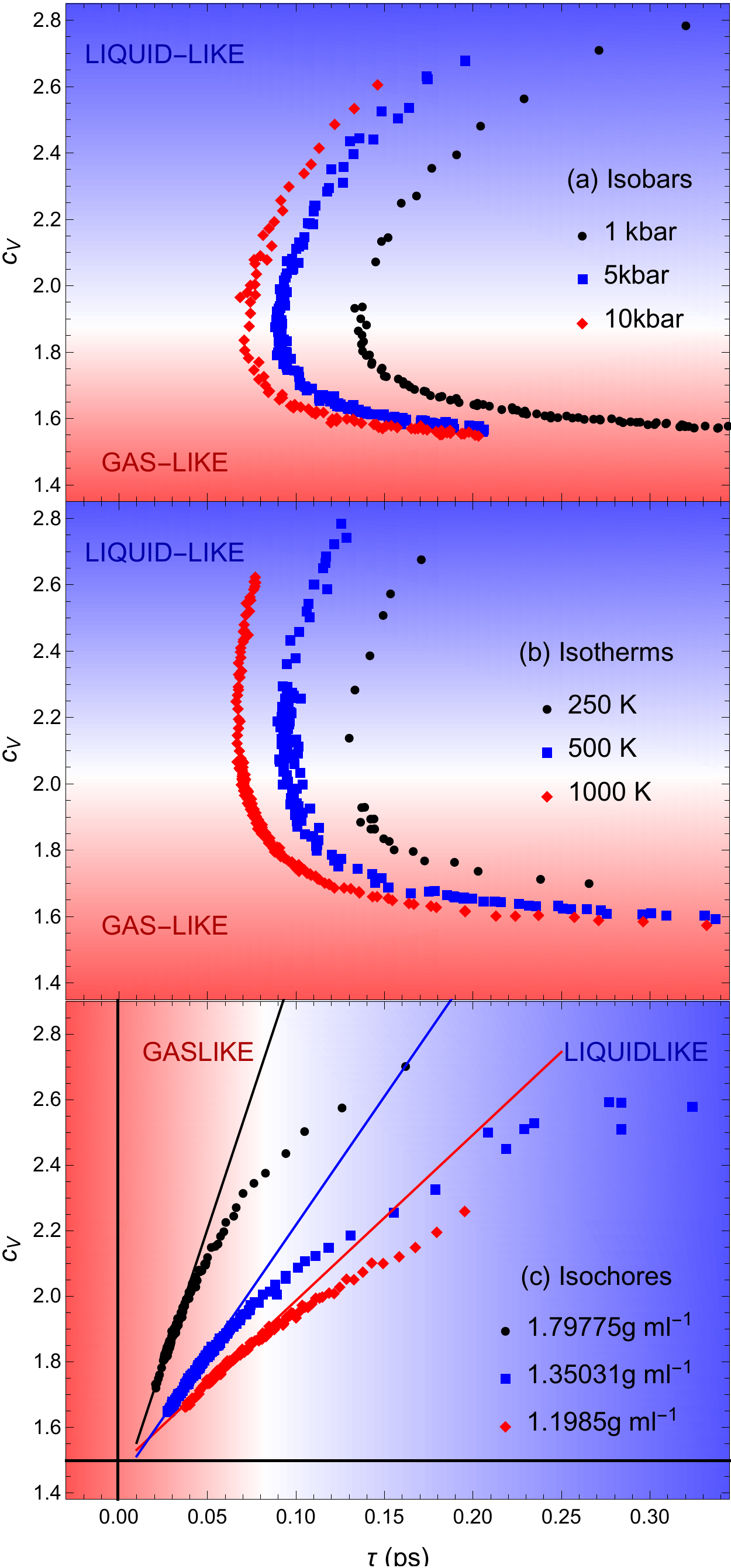}}}
\end{center}
\caption{Specific heat $c_v$ ($k_{\rm B}=1$) as a function of relaxation time $\tau$, calculated along three (a) isobars ; (b) isotherms; (c) isochores. The solid lines in (c) represent the approach of $c_V$ and $\tau$ to their limiting ideal-gas value $c_V=3/2$ at large temperature and are meant as guides for the eye.}
\label{fig:cvtau}
\end{figure}

We next investigate the relationship between dynamics and thermodynamics by plotting $c_V$ as a function of $\tau$ along these nine paths in Fig. \ref{fig:cvtau}. Along isobars and isotherms, $c_V(\tau)$ has clear turning points corresponding to the minima in $\tau$.  This turning point occurs close to $c_V=2$, such that the crossover in $c_V$ from liquidlike to gaslike corresponds to the dynamical transition in $\tau$. Along isochores, the situation is again more subtle. However, since the ideal gas limit as $T \rightarrow \infty$ corresponds to both $c_V \rightarrow 3/2$ and $\tau \rightarrow 0$, the function $c_V(\tau)$ in its gaslike regime must approach $3/2$ as $\tau \rightarrow 0$. Inspection of Fig. \ref{fig:cvtau}c reveals that this is indeed the case, and that $c_V(\tau)$ settles into this limiting behaviour again close to $c_V=2$. 

Notably, Figures \ref{fig:cveta} and \ref{fig:cvtau} show the significant path dependence of $c_V$ on $\eta$ and $\tau$: $c_V$ depends differently on these parameters along isochoric, isobaric and isothermic paths as well as different conditions for each path. Moreover, switching the dependence of $c_V$ from $\eta$ to the related $\tau$ completely changes the shape of the curves.

We now come to the pinnacle of these analyses. The minima of $\tau$ are path-dependent as mentioned earlier. Instead of plotting $c_V(\tau)$, we plot $c_V(c\tau)$ and as the function of the key parameter in the solution of Eq. \eqref{gener3} which we have called the dynamical length $\lambda_d=c\tau$ ($c$ is calculated as $c=\sqrt\frac{G}{\rho}$ as discussed below Eq. \eqref{gener3}). The physical reason for this dependence is provided by Eq. \eqref{kgap}: $c\tau$ governs the propagation of collective modes (phonons) in liquids and supercritical fluids. Since each phonon carries energy ($T$ in the classical harmonic case), $c_V$ is governed by $c\tau$. Moreover, this dependence is predicted to be unique and path-independent, in contrast to the dependence of $c_v$ on $\eta$ or $\tau$ plotted earlier.

\begin{figure*}
        
                \includegraphics[width=0.7\linewidth]{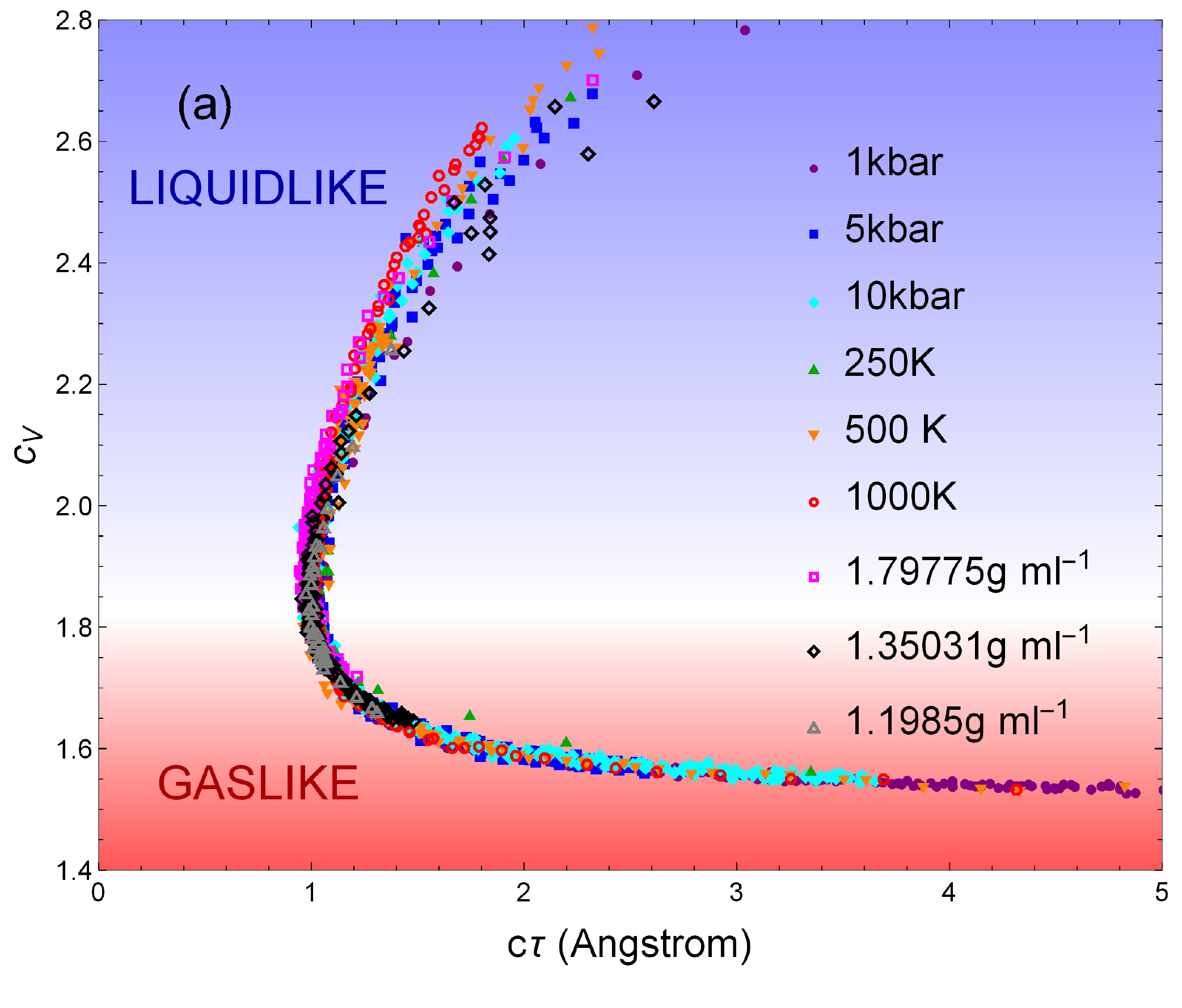}

        \begin{tabular}{ll}
            \includegraphics[width=0.45\linewidth]{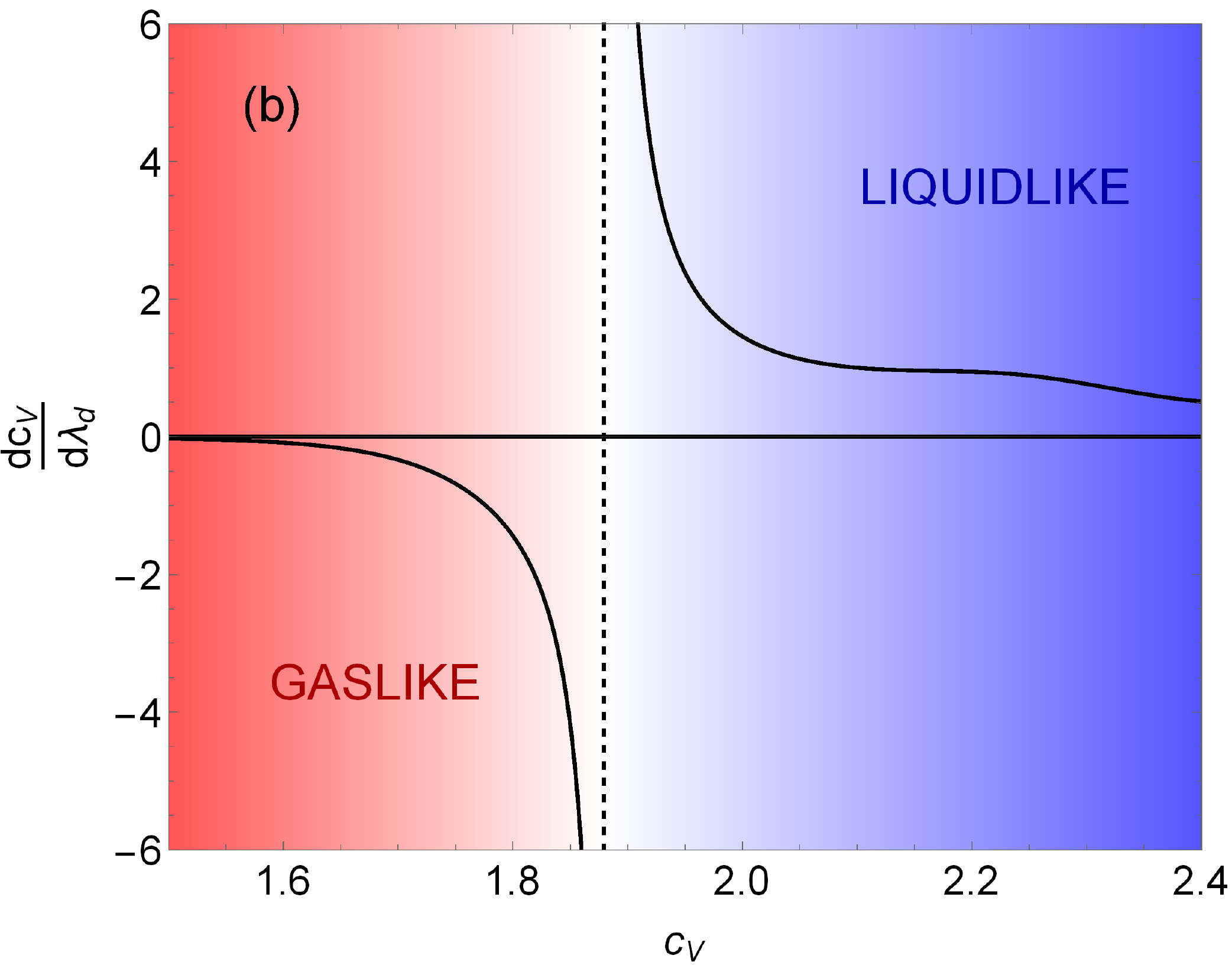}
        \end{tabular}
    \caption{(a) $c_V$ as a function of the dynamical length, $\lambda_d=c\tau$ across 9 paths spanning the supercritical state up to 330 $T_c$ and 8000 $P_c$. $k_{\rm B}=1$. All these paths collapse onto a single curve and undergo a unified dynamic-thermodynamic transition at the path-independent point $c_V = 1.88 $ and $\lambda_d=1~\rm{\AA}$; (b) Divergence in $\frac{\rm{d}c_V}{\rm{d} \lambda_d}$ as a function of $c_V$ at the point $c_V = 1.88$. Curves are obtained by fitting $c_V(\lambda_d)$ data from all nine supercritical paths.}
    \label{fig:cvctau}
\end{figure*}

The result of this master plot in Fig. \ref{fig:cvctau}a is striking. Despite the clear difference among paths of the same and different type seen in Figs. \ref{fig:cveta}, \ref{fig:tau} and \ref{fig:cvtau}, the functions $c_V(c \tau)$ \textit{along all deeply supercritical paths} converge into the same ``c"-shaped curve in Fig. \ref{fig:cvctau}a.

The values of $c_V\approx 2$ and $\lambda_d\approx 1$ \AA\ at the inversion point in Fig. \ref{fig:cvctau}a are physically significant. Brillouin was first to recognize the significance of $c_V=2$ \cite{bril}. The solid energy can be written as the sum of energies of one longitudinal and two transverse modes as $2\times NT/2+2\times(NT/2+NT/2)$, where $NT/2$ is the kinetic and potential energy component of each mode (the two components are equal according to the equipartition theorem), $T$ is the energy of one mode in the harmonic classical case and $N$ is the number of particles. Brillouin assumed that liquids do not support solid-like transverse modes, implying that the potential energy component of these modes, $NT/2$, becomes 0. This gives the energy $2NT$ and $c_V=2$ \cite{bril}. Since this contradicted the experimental $c_V\approx 3$ of liquids at the melting point, a proposition was made that liquids consist of crystallites with easy cleavage directions so that liquids have $c_V=3$ and can flow at the same time. We now understand that (a) liquids do, in fact, support solid-like transverse modes, albeit at high-$k$ only as in Eq. \eqref{kgap} (or at high-frequency for propagating modes $\omega>\frac{1}{\tau}$ \cite{ropp,physrep}) and (b) these modes do disappear eventually, but only well above the melting point and at very high temperature where $\tau$ becomes comparable to Debye vibration period. At this point, $k_g$ in Eq. \eqref{kgap} becomes close to the largest wavevector set by the interatomic separation in the system (ultaviolet, or UV, cutoff), and all transverse waves disappear from the spectrum, resulting in $c_V=2$. This disappearance of transverse modes and the condition $c_V=2$ correspond to the Frenkel line, which separates two dynamical regimes of particle motion: combined oscillatory and diffusive motion in the low-temperature ``rigid'' liquid (rigid in a sense of its ability to support solid-like transverse modes); and purely diffusive motion in the high-temperature non-rigid gaslike fluid \cite{ropp}. The loss of the oscillatory component at the Frenkel line transition gives the second, dynamical, criterion of the line: the disappearance of the minima of the velocity autocorrelation function (VAF). The thermodynamic criterion $c_V=2$ and the dynamical VAF criterion give coinciding lines on the phase diagram \cite{vaf}.

A change in dynamics implies and is implied by a change in structure \cite{ropp}, and, stimulated by the Frenkel line idea, the transitions at the Frenkel line have been subsequently experimentally observed in supercritical Ne \cite{f1}, CH$_4$ \cite{f2}, N$_2$ \cite{f3}, C$_2$H$_6$ \cite{f4} and CO$_2$ \cite{f5} using X-ray, neutron and Raman scattering techniques. A crossover in structure has been seen across the FL in simulated argon \cite{Wang2017}, using the same potential as in this work, however this crossover is subtle as compared to the transition presented here.

As mentioned earlier, $c_V=2$ corresponds to the purely harmonic case where the mode energy is $T$. Anharmonicity can change this result by a relatively small amount \cite{ropp}, and the disappearance of transverse modes corresponds to $c_V=2$ approximately. We also note that similarly to solids, plane waves decay in liquids. The decay mechanisms in solids include anharmonicity, defects and structural disorder present in, for example, glasses. Despite this decay, high-temperature specific heat in solids is governed by phonons. In liquids, the additional decay mechanism is related to atomic jumps \cite{ropp}. Nevertheless, the propagation length of high-frequency excitations in liquids and supercritical fluids is on the order of nanometers, as evidenced by experiments and modelling \cite{hoso1,hoso2,hoso3,prl}. This is similar to room-temperature solids where the lifetime of high-frequency phonons is on the order of picoseconds and the propagation range is on the order of nanometers \cite{Jain2016} and where disorder and/or defects reduce these values further. This is also similar to glasses which are structurally similar to liquids \cite{ruffle}. Therefore, phonon excitations govern the specific heat in liquids and supercritical fluids to the same extent they do in solids. 

The significance of the propagation range $\lambda_d$ reaching about 1 \AA\ at the inversion point in Fig. \ref{fig:cvctau}a is that it corresponds to the shortest distance in the system (UV cutoff), the interatomic separation on the order of Angstroms in condensed matter phases. This distance is fixed by fundamental physical constants in the form of the Bohr radius which, together with the Rydberg energy, sets the viscosity minimum \cite{sciadv}. The UV cutoff puts the upper limit for $k$-points for propagating waves in the system. Recall that $k_g$ in \eqref{kgap} sets the range of transverse waves. When $k_g$ becomes comparable to the largest, Debye, wavevector, all transverse modes disappear from the system spectrum. We have calculated the Debye wavevector $k_{\rm D}=(6\pi^2 n)^\frac{1}{3}$ \cite{landau}, where $n$ is particle density, to be about 1.0 \AA$^{-1}$ at the inversion point. Equating it to $\frac{1}{2c\tau}$ in \eqref{kgap} gives $\lambda_d=c\tau\approx$0.5 \AA. This is consistent with about 1 \AA\ at the inversion point in Fig. \ref{fig:cvctau}a, given the approximate nature of Debye model and that \eqref{kgap} applies to the linear part of the dispersion relation only but not to the range where $\omega(k)$ flattens off close to the zone boundary at $k=k_{\rm D}$. 

In the gaslike state, the quantity $c\tau$ increases towards the ideal limit of $c_V = 3/2$ in Fig. \ref{fig:cvctau}a. In the gaslike state, $c\tau$ is related to the particle mean free path since the speed of sound $c$ is proportional to the thermal velocity $v_{\rm{th}}$ and $\tau$ is related to the average time between particle collisions \cite{frenkel}. The decrease of heat capacity from 2 to 3/2 in the gaslike regime of fluid dynamics can be explained in terms of longitudinal modes at short distance in the fluid vanishing as $c\tau$ exceeds their wavelength \cite{ropp}. The collapse of all curves in the gaslike regime in Fig. \ref{fig:cvctau}a implies that $c\tau$ is the \textit{only} parameter necessary to characterise the loss of these degrees of freedom in this regime.

We reiterate to make explicit the significance of this effect in Fig. \ref{fig:cvctau}a : paths in the supercritical region, separated by orders of magnitude in temperature and pressure, all sport the same $c_V$ vs $c \tau$ curves. The heat capacity, usually defined as a function $c_V(T, P)$ of both temperature and pressure, is here reduced to a near-universal function of the single dynamical fluid elasticity length $\lambda_d=c\tau$ which sets the range of propagation of solid-like transverse modes as discussed above. What's more, the apex, or the inversion point, of this unified curve takes place at $c_V \approx 1.9$, close to the key value $c_V=2$ at the Frenkel line and to the UV cutoff length of the system.

We plot temperature and pressure points at which $c_V=1.9$ at the inversion point on all paths as stars in Fig. \ref{fig:phasediagram}. We observe that the location of these points varies by orders of magnitude on the phase diagram, yet they all coincide on our master plot in Fig. \ref{fig:cvctau}a.

This curve $c_V(c \tau)$ therefore represents a \textit{dynamically driven transition} between liquidlike and gaslike states which is present across the supercritical region and is \textit{path independent}. We call this transition a ``c"-transition due to the shape of Fig. \ref{fig:cvctau}a. The inversion point of all coinciding curves in Fig. \ref{fig:cvctau}a is well-defined and therefore serves as an unambiguous transition point between liquidlike and gaslike states in the supercritical region. The inversion point does not depend on a theory such as that underlying the thermodynamic criterion of the Frenkel line $c_V=2$ and the dynamical VAF criterion. 

The transition is further illustrated in Fig. \ref{fig:cvctau}b: we fit the data in Fig. \ref{fig:cvctau}a, calculate the derivative  $\frac{\rm{d}c_V}{\rm{d}\lambda_d}$ and plot the derivative vs $c_V$, observing the divergence at $c_V$ close to the key value $c_V=2$ at the Frenkel line. 

\begin{figure}
             \includegraphics[width=0.95\linewidth]{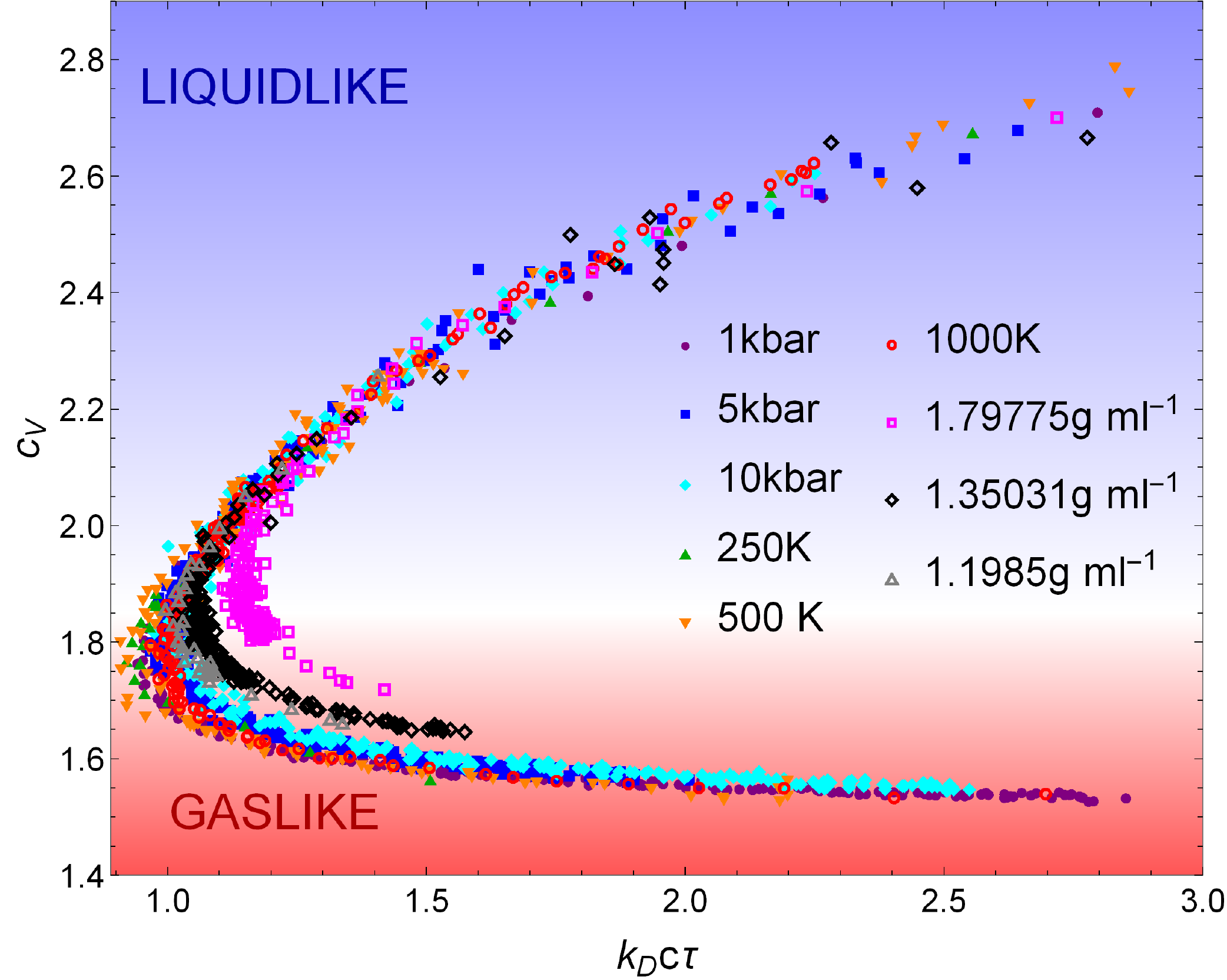}
    \caption{$c_V$ as a function of $c \tau k_{\rm{D}}$, where the curves similarly converge in the liquidlike state at the path-independent point close to $c_V=2$.}
    \label{fig:cvctau1}
\end{figure}

An alternative way to collapse the data from different paths is informed by the theory of liquid thermodynamics. In this theory, the liquid energy at low temperatures, where transverse waves exist below the Frenkel line, depends on the ratio of $k_g$ in \eqref{kgap} and the Debye wavector $k_{\rm D}$: $\frac{k_g}{k_{\rm{D}}}$ \cite{ropp,prl,jpcm} (or, alternatively, on $\frac{\omega_{\rm F}}{\omega_{\rm{D}}}$ for propagating waves \cite{ropp}, where $\omega_{\rm F}=\frac{1}{\tau}$ and $\omega_{\rm{D}}$ is Debye frequency). The physical picture here is that shorter $\tau$ at high temperature results in progressive disappearance of transverse waves from the system spectrum \cite{ropp} (this theory has undergone a detailed and rigorous test \cite{fluids} and explains the decrease of $c_V$ with temperature in Fig. \ref{fig:totalcompare} (b)). However, this disappearance of transverse modes can only proceed up to the largest $k$-point, $k_{\rm{D}}$, set by the UV cutoff in the system as discussed earlier. As a result, $\frac{k_g}{k_{\rm{D}}}$ enters the equation for the liquid energy. We calculate $k_{\rm{D}}$ as $(6\pi^2 n)^\frac{1}{3}$ \cite{landau} and plot $c_V$ as a function of dimensionless product $c\tau k_{\rm{D}}$ in Fig. \ref{fig:cvctau1}. We observe that the curves from all paths collapse in the region of large $c_V$ down to about $c_V=2$, showing that $c_V$ below the Frenkel line depends on the product $c \tau  k_{\rm{D}}$ only, as predicted theoretically. This is followed by the divergence of $c_V$ below $c_V=2$ in the gaslike state along different paths, presumably because the solid-like concepts underlying $k_{\rm{D}}$ become progressively less relevant in the gaslike state at high temperature.

The key result of this work is visible in both scaling graphs in Figs. \ref{fig:cvctau}a and \ref{fig:cvctau1}: the data collapse in either large or small-value range of $c_V$; the curves diverge from each other in the other range; and this divergence takes place near the key value $c_V=2$.

\begin{figure}
             \includegraphics[width=0.95\linewidth]{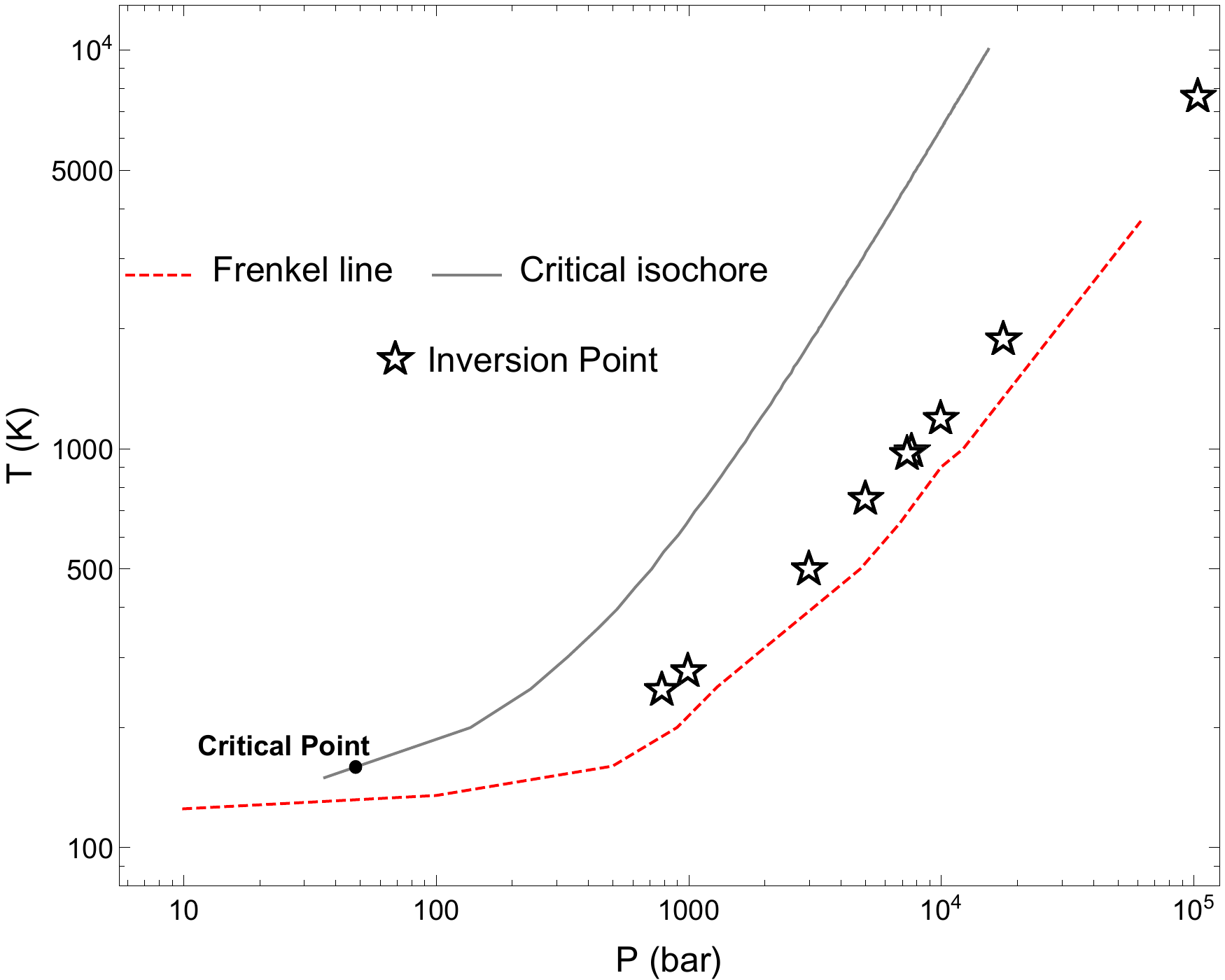}
    \caption{Phase diagram showing the inversion points of the ``c"-transition corresponding to $c_V \approx 1.9$ in Fig. \ref{fig:cvctau}, the Frenkel line determined by the VAF criterion \cite{vaf}, the critical point, and the critical isochore.}
    \label{fig:phasediagram2}
\end{figure}

As mentioned earlier, the dynamical VAF criterion of the Frenkel line corresponds to the disappearance of the oscillatory component of particle motion. In Fig. \ref{fig:phasediagram2}, we plot the line calculated using the VAF criterion, together with the ``c''-transition line determined by ($P$,$T$) at the inversion point where $c_V\approx 1.9$ in Fig. \ref{fig:cvctau}. We also plot the critical isochore for comparison. The Frenkel line from the VAF criterion and the inversion point are close and run parallel to each other. This serves as self-consistency check for our theory and implies that the inversion point can serve as a hallmark and a definition of the supercritical transition between the liquidlike and gaslike states at the FL. As mentioned earlier, the inversion point is unambiguously defined in Fig. \ref{fig:cvctau} and does not depend on a theory such as that underlying the thermodynamic criterion $c_V=2$ and the dynamical VAF criterion.

We note that near the critical point, such as along the 100 bar isobar, the dynamical and thermodynamical properties are strongly affected by near-critical anomalies \cite{deben,stanley}, and the function $c_V(c\tau)$ is affected as a result. The function $c_V(c \tau)$ calculated along this path does not collapse onto the inversion point. On the other hand, our master curve in Fig. \ref{fig:cvctau} is deeply supercritical and is therefore free of the near-critical anomalies. 

The collapse of all curves up to the key value of about $c_V=2$ and divergence of curves along different paths beyond this value has two further implications. First, it suggests that the inversion point $c_V = 2$ is a special point on the phase diagram. Second, if a thermodynamic property has a wide crossover, the behavior of different properties strongly depends on the path taken on the phase diagram. On the other hand, the observed collapse of all paths at the special inversion point close to $c_V=2$ and $\lambda_d = 1$ \AA\ indicates either a sharp crossover or a dynamically driven phase transition related to the ``c"-transition between liquidlike and gaslike states. By sharp we mean more abrupt that previous transitions observed over the FL (see, e.g., Ref. \cite{f1}), possibly involving discontinuities in higher-order thermodynamic derivatives. Within the uncertainty set by fluctuations in our simulations, we do not observe an anomaly of $c_V$ in Fig. \ref{fig:totalcompare}b at temperatures and pressures corresponding to the inversion point (our simulations put an upper boundary of about 0.05$k_{\rm B}$ on the value of a possible anomaly of $c_V$). This does not exclude a weak thermodynamic phase transition, similar to a percolation transition, or a higher-order phase transition seen in higher derivatives of thermodynamic functions. 

In summary, we have discovered a universal, striking, and demonstrative inter-relation between dynamics and thermodynamics using the specific heat $c_V$ and the dynamical parameter of the fluid elasticity length, $ \lambda_d = c \tau$. This connection provides a clear and path-independent transition between liquidlike and gaslike supercritical states, which we call a ``c''-transition. Our ``c"-shaped master curve provides an unambiguous and path-independent criterion for the separation of liquidlike and gaslike states, calculated from accessible quantities in molecular dynamics simulations. The collapse onto this master curve occurs in the supercritical state up to $T = 330 T_c$ and $ P = 8000 P_c$, meaning the transition, and the distinct states it separates, exist over a far larger range of temperatures and pressures than the boiling line which separates subcritical liquids and gases. The collapse is indicative of either a sharp crossover or a new phase transition operating in the supercritical state.

\section{Acknowledgements}

We are grateful to EPSRC and Queen Mary University of London for support and J. C. Dyre for discussions. This research utilized Queen Mary’s MidPlus computational facilities, supported by QMUL Research-IT.


\begin{thebibliography}{9}

\bibitem{landau} L. D. Landau and E. M. Lifshitz, Statistical Physics (Pergamon Press, 1969).

\bibitem{deben} E. Kiran, P. G. Debenedetti and C. J. Peters, Supercritical Fluids: Fundamentals and Applications (NATO Science Series E: Applied Sciences vol 366) (Boston: Kluwer, 2000).

\bibitem{nist} National Institute of Standards and Technology database, see https://webbook.nist.gov/chemistry/fluid.

\bibitem{viscositysize} I.-C. Yeh and G. Hummer, J. Phys. Chem. \textbf{108}, 15873-15879 (2004)

\bibitem{Frenkel2001} D. Frenkel and B. Smit, \textit{Understanding Molecular Simulation:From Algorithms to Applications}, Academic Press, San Diego, San Francisco, (2001).

\bibitem{Balucani1994} U. Balucani and M. Zoppi, \textit{Dynamics of the Liquid State}, Clarendon, Oxford (1994).

\bibitem{Zwanzig1965} R. Zwanzig and R. D. Mountain, J. Chem. Phys. \textbf{43}, 4464 (1965).

\bibitem{Zhang2015} Y. Zhang, A. Otani, E. J. Maginn, J. Chem. Theory Comput. \textbf{11}, 3537 (2015).

\bibitem{sciadv} K. Trachenko and V. V. Brazhkin, Science Adv. {\bf 6}, eaba3747 (2020).

\bibitem{abramson} E. H. Abramson, High Press. Res. \textbf{31}, 544-548 (2011).

\bibitem{excess} I. H. Bell, R. Messerly, M. Thol, L. Costigliola and J. C. Dyre, J. Phys. Chem. B {\bf 123}, 6345 (2019).

\bibitem{maxwell} J. C. Maxwell, Philos. Trans. R. Soc. London 157, 49 (1867).

\bibitem{frenkel} J. Frenkel, Kinetic Theory of Liquids, (New York: Dover, 1955).

\bibitem{ropp} K. Trachenko and V. V. Brazhkin, Rep. Prog. Phys. \textbf{79}, 016502 (2016).

\bibitem{physrep} M. Baggioli, M. Vasin, V.  V. Brazhkin and K. Trachenko, Physics Rep. {\bf 865}, 1 (2020).

\bibitem{prl} C. Yang, M. T. Dove, V. V. Brazhkin, K. Trachenko, Phys. Rev. Lett. {\bf 118}, 215502 (2017). 

\bibitem{jac} B. Jakobsen, T. Hecksher, T. Christensen, N. Boye Olsen, J. C. Dyre, K. Niss, J. Chem. Phys. \textbf{136}, 081102 (2012).

\bibitem{iwa} T. Iwashita, D. M. Nicholson, T. Egami, Phys. Rev. Lett. \textbf{110}, 205504 (2013).

\bibitem{chapman} S. Chapman and T. G. Cowling, The mathematical theory of non-uniform gases (Cambridge University Press, 1990). 

\bibitem{zaccone} A. Zaccone, J. Phys.: Condens. Matter {\bf 32}, 203001 (2020). 

\bibitem{dyre} J. C. Dyre, Rev. Mod. Phys. {\bf 78}, 953 (2006).

\bibitem{bril} L.  Brillouin, Tensors in Mechanics and Elasticity (Academic, New York, 1964).

\bibitem{f1} C. Prescher, Y. D. Fomin, V. B. Prakapenka, J. Stefanski, K. Trachenko, V. V. Brazhkin, Phys. Rev. B {\bf 95}, 134114 (2017).

\bibitem{f2} D. Smith, M. A. Hakeem, P Parisiades, H. E. Maynard-Casely, D. Foster, D. Eden, D. J. Bull, A. R. L. Marshall, A. M. Adawi, R. Howie, A. Sapelkin, V. V. Brazhkin, J. E. Proctor, Phys. Rev. E {\bf 96}, 052113 (2017).

\bibitem{f3} J. Proctor, C. G. Pruteanu, I. Morrison, I. F. Crowe, and J. S. Loveday, J. Phys. Chem. Lett. {\bf 10}, 6584 (2019).

\bibitem{f4} J. Proctor, M. Bailey, I. Morrison, M. A. Hakeem, I. F. Crowe, J. Phys. Chem. B {\bf 122}, 10172 (2018).

\bibitem{f5} C. J. Cockrell, O. Dicks, L. Wang, K. Trachenko, A. K. Soper, V. V. Brazhkin, S. Marinakis, Phys. Rev. E  {\bf 101}, 052109 (2020).

\bibitem{Wang2017} L. Wang, C. Yang, M. T. Dove, Y. D. Fomin, V. V. Brazhkin, K. Trachenko, Phys. Rev. E. \textbf{95}, 032116 (2017).

\bibitem{jpcm} Y. Fomin et al, J. Phys.: Condens. Matt. {\bf 30}, 134003 (2018).

\bibitem{hoso1} S. Hosokawa et al, Phys. Rev. Lett. {\bf 102}, 105502 (2009).

\bibitem{hoso2} S. Hosokawa et al, J. Phys. Condens. Matt. {\bf 25},  112101 (2013).

\bibitem{hoso3} S. Hosokawa et al, J. Phys. Condens. Matt. {\bf 27}, 194104 (2015).

\bibitem{Jain2016} A. Jain and A. J. H. McGaughey, Phys. Rev. B {\bf 93}, 081206 (2016).

\bibitem{ruffle} B. Ruffl\'e, G. Guimbretie\'re,  E. Courtens,  R. Vacher and G. Monaco, Phys. Rev. Lett. {\bf 96}, 045502 (2006).

\bibitem{fluids} J. Proctor, Physics of Fluids, {\bf 32}, 107105 (2020).

\bibitem{stanley} L. Xu et al, PNAS {\bf 102}, 16558 (2005). 

\bibitem{vaf} V. V. Brazhkin, Yu. D. Fomin, A. G. Lyapin, V. N. Ryzhov, E. N. Tsiok and K. Trachenko, Phys. Rev. Lett. {\bf 111}, 145901 (2013).

\end{thebibliography}
\end{document}